\documentclass[a4paper,12pt]{article}
\usepackage[utf8]{inputenc}
\usepackage{cancel}
\usepackage{ulem}
\usepackage{amsfonts}
\usepackage{amssymb}
\usepackage{graphicx}
\usepackage{amsmath}
\usepackage{enumerate}
\usepackage{mathtools}
\usepackage{subfig}
\usepackage{color}
\usepackage{tikz}
\usepackage{float}
\usepackage{here}
\usepackage{cite}
\usepackage{mathrsfs}
\usepackage{float,epsfig}
\usepackage{dcolumn}
\usepackage{lscape}
\usepackage{subfig}
\usepackage{graphicx}
\usepackage{bm}
\usepackage{amsmath,amssymb,amsthm}
\usepackage[colorlinks=true,linkcolor=blue,citecolor=red]{hyperref}
\usepackage{multirow}

\usetikzlibrary{calc}
\newcommand{\be}{\begin{equation}}
\newcommand{\ee}{\end{equation}}
\newcommand{\bea}{\setlength\arraycolsep{2pt} \begin{eqnarray}}
\newcommand{\eea}{\end{eqnarray}}

\def\0{{\sst{(0)}}}
\def\1{{\sst{(1)}}}
\def\2{{\sst{(2)}}}
\def\3{{\sst{(3)}}}
\def\4{{\sst{(4)}}}
\def\5{{\sst{(5)}}}
\def\6{{\sst{(6)}}}
\def\7{{\sst{(7)}}}
\def\8{{\sst{(8)}}}
\def\sst#1{{\scriptscriptstyle #1}}

\makeatletter \@addtoreset{equation}{section}

\begin{document}

\title{\textbf{Stability and Criticality Behaviors of Accelerating Charged AdS Black Holes in Rainbow Gravity}}
\author{ Hajar Belmahi\footnote{hajar\_belmahi@um5.ac.ma},  Maryem Jemri, Rajaa Salih\thanks{
Authors are listed in alphabetical order.} \\
{\small ESMaR, Faculty of Science, Mohammed V University in Rabat, Rabat, Morocco}}
\maketitle

\begin{abstract}
In this work, we   investigate the thermodynamical properties  of accelerated charged
Anti-de Sitter black holes in the context of rainbow gravity.  Concretely,   we compute  the   corresponding quantities needed to  study  the  thermal stability and   the critical behaviors including the phase transitions.  Linking the  acceleration parameter $A$ and the horizon radius $r_h$ via a constant parameter $a$,  we discuss 
the $P$-$v$ criticality behaviors  by  calculating  the critical pressure
$P_c$,  the critical temperature $T_c$ and the  critical specific volume $v_c$ in terms of  $a$ and  the rainbow gravity parameter  $\varepsilon$.  As results, we reveal  that  the ratio $ \dfrac{P_{c}v_{c}}{T_{c}}$ is an universal number with respect to the charge.  In small limits of the external parameters,  we recover    the Van der Waals fluid    behaviors.  After that,  we examine the Joule-Thomson expansion effects  for  such black holes.   We observe that the  similarities and  the differences with  Van der Waals fluids depend on the region parameters.

{\noindent} 

\end{abstract}

%

\newpage


\newpage

\section{Introduction}

In the context of general relativity, black holes are considered as  solutions to the Einstein field equations \cite{1,102}. 
Revolutionary observations have confirmed their existence.
The first definitive screening of a stellar-mass black hole, Cygnus X-1, was accomplished through
the Isaac Newton Telescope (INT) \cite{2}, revealing a compact object with mass exceeding the
Tolman-Oppenheimer-Volkoff limit for neutron stars.Years later,  the Laser Interferometer
Gravitational-Wave Observatory (LIGO) opened a new era by directly observing gravitational
waves from merging black holes \cite{3}, further validating their dynamic role in spacetime. Recently, the Event Horizon Telescope (EHT)  furnished the first-ever image of an shadow of black holes, providing a  visually direct evidence of event horizon \cite{4,401,402}.   In the strong-field regime, this has been considered as  confirming key predictions of general relativity\cite{1}. These
discoveries have lighted intense research into black hole physics, particularly regarding their
impact on the  spacetime geometries and their potential as probes for alternative theories of gravity.

Physical studies of black holes have evolved to treat them as systems with thermodynamical properties \cite{6,71,72,73,74,75,76,77,78,79,80,81,82,83,Karch}, whose laws are analogous to those of classical thermodynamic systems. Further research incorporating quantum effects led to Hawking's theoretical prediction of the black hole radiation, which reveals that black holes possess a temperature proportional to their surface gravity \cite{7}, as well as entropy \cite{8}. These advances have revolutionized our perception of the laws of physics, the nature of space-time and the fundamental principles governing the universe.
Since the discovery of the thermodynamical properties of black holes, intense research
has been devoted to the study of various types of black holes in different spacetime geometries, such as Sitter and Anti-de Sitter (AdS) spaces. These studies have shown that the presence of the cosmological constant influences the thermal stability of black holes \cite{9,m1,m3,99,999,EslamPanah:2025zcm,Mancilla}. In
what concerns black holes in AdS spacetime, they can undergo phase transitions \cite{10}, notably the Hawking-Page phase transition
and the small/large black hole phase transition \cite{9}.

 Recently, a special emphasis has been put on the investigation of   accelerating  AdS  black holes by considering the thermodynamics  behaviors \cite{11,111}. Precisely, 
  the $P-v$ criticality and the phase transition of charged  accelerating AdS black holes in the extended thermodynamic phase space  have been studied\cite{12,m2}.
  
In the AdS context, the C-metric in rainbow gravity offers an effective framework to describe uniformly accelerating black holes embedded in  AdS backgrounds, modified by energy-dependent deformations of the spacetime geometry \cite{13}. Motivated by quantum gravity approaches, the rainbow gravity introduces modifications to the standard dispersion relations, leading to a deformation of the metric felt by test particles. Combining rainbow gravity with C-metric describing  acceleration, this framework allows one to explore novel thermodynamic phenomena induced by both acceleration and high-energy corrections \cite{14,EslamPanah}.

The aim of this work is to explore the thermodynamical properties  of accelerated charged
AdS black holes in the context of rainbow gravity. Precisely,   we investigate  the  thermal 
stability and   the critical behaviors including the phase transitions by calculating   the associated  quantities.  Linking the  acceleration parameter $A$ and the horizon radius $r_h$ via a constant parameter $a$,  we approach
the $P-v$ criticality  by computing  the critical pressure
$P_c$,  the critical temperature $T_c$ and the  critical specific volume $r_c$ in terms of  $a$ and  the rainbow gravity parameter $\varepsilon$. We show  that  the ratio $ \dfrac{P_{c}r_{c}}{T_{c}}$
is a universal number with respect to the charge. In small limits of the external parameters,  we recover    the Van der Waals fluid    behaviors.  After that,  we  inspect  the Joule-Thomson expansion effects  for  such black holes.   We  show  that the  similarities and  the differences with  Van der Waals fluids depend on the region parameters.

The  organization of this work  is as follows.  In section 2,   we present a concise discussion on   the   accelerating charged AdS black holes in
 rainbow  gravity.
In section 3, we  calculate   and examine   the  thermodynamical quantities associated with such a 
solution in order to  study the
stability behaviors.  In  section 4, we   investigate   the
 $P-v$   criticality,  the phase transitions, and   Joule-Thomson expansion effects. The last section   is devoted to concluding remarks.

\section{Charged accelerating  black holes  in
 rainbow  gravity}
\begin{figure}
    \centering
    \begin{tabular}{cc}
       \includegraphics[width=7cm,height=7cm]{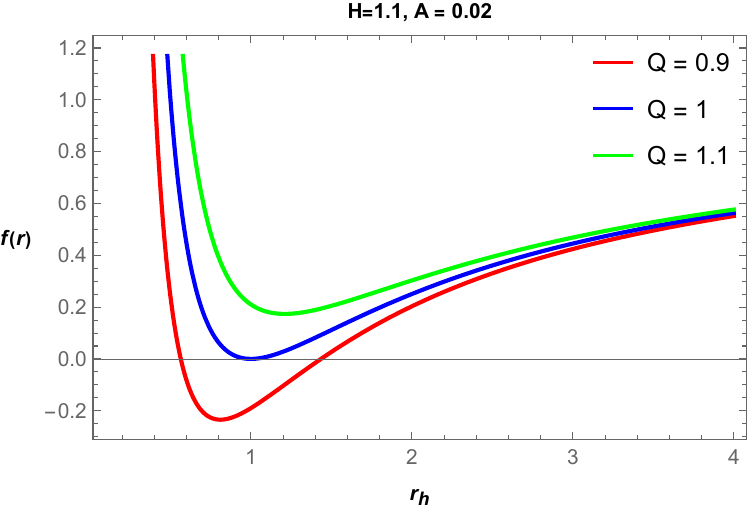} & 
         \includegraphics[width=7cm,height=7cm]{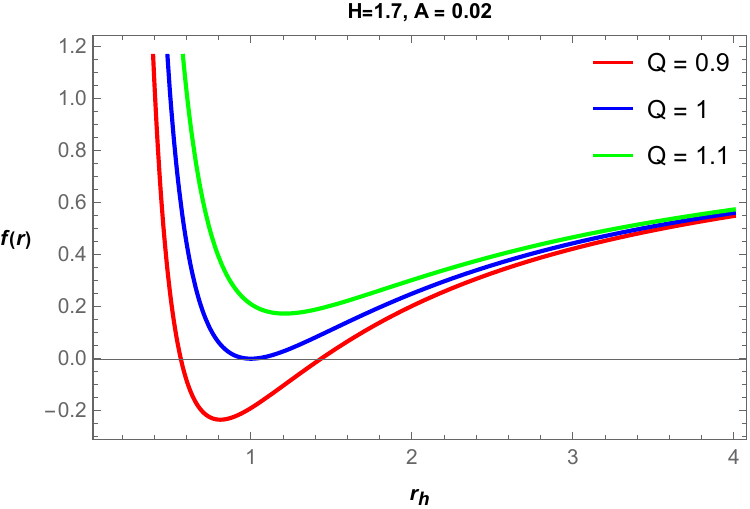} 
    \end{tabular}
     \caption{Effect of the charge parameter $Q$ and the rainbow function $H(\varepsilon)$ on the metric function $f(r)$.}
    \label{Fig3.1}
\end{figure}

 In this section, we present a concise discussion   on charged  accelerating black holes  in the rainbow  gravity.
By considering the C-metric introduced in \cite{14} and incorporating the energy-dependent modifications discussed in \cite{15}, one assumes  the charged, the  energy-dependent C-metric given by
\begin{equation}
ds^2 = \frac{1}{\omega^2} \left( -\frac{f(r)}{F^2(\varepsilon)} dt^2 + \frac{dr^2}{f(r) H^2(\varepsilon)} + \frac{r^2}{H^2(\varepsilon)} \left( \frac{d\theta^2}{g(\theta)} + \frac{g(\theta) \sin^2\theta}{K^2} d\phi^2 \right) \right), 
\end{equation}
where one has
\begin{eqnarray}
f(r) &= &(1 - A^2 r^2) \left( 1 - \frac{2M}{r} + \frac{Q^2}{r^2} \right) - \frac{\Lambda r^2}{3 H^2(\varepsilon)} \\
g(\theta) &=& 1 + 2 M A \cos\theta + Q^2 A^2 \cos^2\theta,    
\end{eqnarray}
and the conformal factor is given by
\begin{equation}
\omega(r, \theta) = 1 + A r \cos\theta. 
\end{equation}
This factor  can determine  the conformal infinity (boundary) of the AdS spacetime.  $H(\varepsilon)$ and $F(\varepsilon)$ are known by   rainbow functions  depending on $\varepsilon$ being expressed  as  follows
\begin{equation}
H(/var\epsilon)=  \sqrt{1-\gamma \varepsilon^2} \qquad F(\varepsilon)=1,
\end{equation}
where one has used  $\varepsilon= \frac{E}{E_{p\ell}}$  where $E$ and  $E_{p\ell} $   represent the energy of the system and the Planck energy, respectively.
It denoted that $\gamma$ is a dimensionless free parameter  which will be fixed to one for simplicity reasons.  The parameters $\Lambda$, $M$, and $Q$ correspond to the cosmological constant, the total mass, and the electric charge of the black hole, respectively.  The acceleration parameter $A > 0$ describes the magnitude of the black hole acceleration. Throughout this work, we adopt natural units with $G = c = 1$.
The parameter $K$ is associated with the presence of a cosmic string \cite{15}. It is determined by analyzing the angular part of the metric and the behavior of the function $g(\theta)$ at the poles, $\theta_+ = 0$ and $\theta_- = \pi$. Regularity of the metric at a given pole requires
\begin{equation}
K_{\pm} = g(\theta_{\pm}) = 1 \pm 2MA + Q^2 A^2.  
\end{equation}
Before approaching  the thermodynamical behaviors of such solutions,  we first examine  the effect of the charge parameter \( Q \) on the horizon structure of the black holes.  Indeed,  this  is shown in Fig.(\ref{Fig3.1}). It is evident that increasing the charge alters the behavior of the metric function \( f(r) \). The black hole exhibits three configurations depending on a critical charge values $Q_c$ providing  a double zero of $f(r)=0$. They are two  horizons (the inner and outer horizons) and naked singularity.    From this figure it has been remarked that $Q>Q_{c}$,  one has  a naked
singularity. For $
Q=Q_{c}$ and $Q<Q_{s}$, however, two types of solution are possible
corresponding respectively to extremal  and non-extremal  black holes. As \( Q \) increases, these horizons approach each other and eventually merge. This disappearance of the horizons indicates a transition to a naked singularity.
We also observe that a higher value of the rainbow function $H(\varepsilon)$ influences the behavior of the metric function, as shown in  Fig.(\ref{Fig3.1}). Specifically, increasing $H(\varepsilon)$ leads to a slight shift of the horizon to a larger $r$, and in the large-$r$ region, the values of $f(r)$ tend to decrease. This modification in the spacetime geometry may affect the thermodynamic stability and the phase structure of the black hole. The analysis also reveals that increasing the acceleration parameter results in the emergence of two horizons, more massive black holes possess larger radii and increasing the magnitude of the cosmological constant \( |\Lambda| \) tends to produce smaller black holes.

\section{Thermal and stability behaviors of charged accelerating black holes}
This section is devoted to the computations of relevant thermodynamical quantities  needed to approach certain behaviors including the stability.    Precisely, we calculate  thermal   stability quantities including the the temperature and the heat capacity. 
\subsection{Thermal behaviors }
To start, one needs to calculate   the mass quantity.  In particular,  we obtain the mass as a function of the horizon radius \( r_h \) by  imposing the constraint \( f(r) = 0 \).  Indeed, the mass  is found to be 
\begin{equation}
M = \frac{ \left( \Lambda + 3A^2 H^2(\varepsilon) \right) r_h^4 + 3H^2(\varepsilon) \left( A^2 Q^2 - 1 \right) r_h^2 - 3Q^2 H^2(\varepsilon) }{ 6r_h \left( A^2 r_h^2 - 1 \right) H^2(\varepsilon) }.    
\end{equation}
To get the Hawking temperature, we use the relation \( T_H = \kappa / (2\pi) \), where the surface gravity \( \kappa \) is defined as
\begin{equation}
\kappa = \left. \frac{ \mathrm{d}f(r) }{ \mathrm{d}r } \right|_{r = r_h} \frac{H(\varepsilon)}{2F(\varepsilon)}.    
\end{equation}
This leads to the following expression for the Hawking temperature
\begin{equation}
T_H = \frac{ \Lambda r_h^2 (3 - A^2 r_h^2) + 3H^2(\varepsilon) \left( \frac{Q^2}{r_h^2} - 1 \right) \left( A^2 r_h^2 - 1 \right)^2 }{ 12\pi r_h H(\varepsilon) \left( A^2 r_h^2 - 1 \right) F(\varepsilon) }.  \label{xxx} 
\end{equation}

 \begin{figure}[h!]
    \centering
    \begin{tabular}{cc}
       \includegraphics[width=7.5cm,height=7cm]{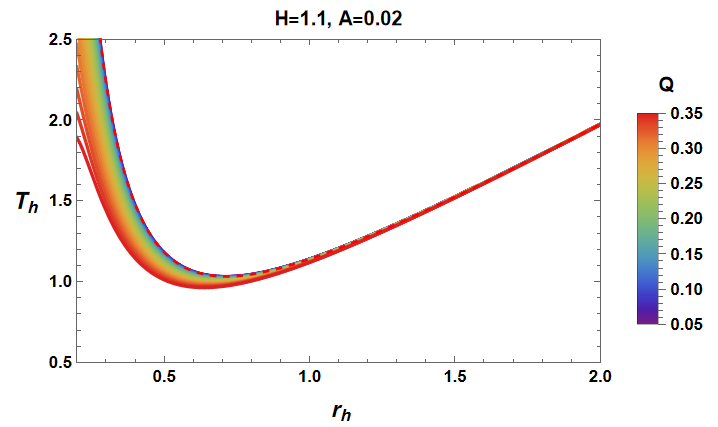} & 
         \includegraphics[width=7.5cm,height=7cm]{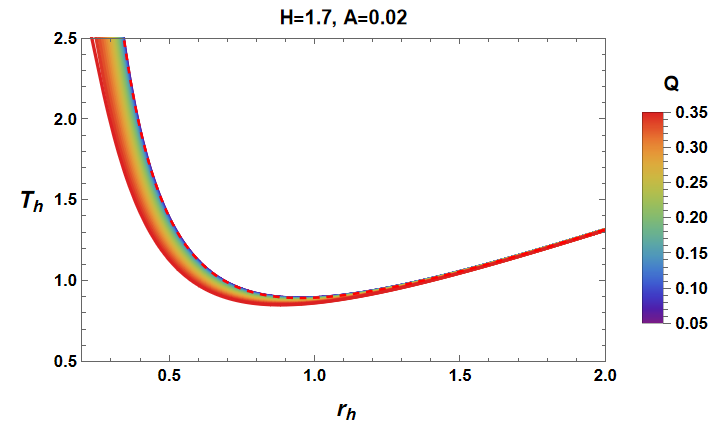} 
    \end{tabular}
    \caption{Effect of the charge parameter $Q$ and the rainbow function $H(\varepsilon)$ on the Hawking temperature $T_H$.}
    \label{Fig3.2}
\end{figure}

It is worth noting that setting the charge to zero yields the temperature of an accelerating AdS black hole in rainbow gravity \cite{EslamPanah:2024dfq}. To inspect the associated thermal behaviors,  we plot the temperature against the radius of event horizon in Fig.(\ref{Fig3.2}) for different values of the charge parameter $Q$ and  the rainbow function $H(\varepsilon)$.  As the charge increases, the temperature curve dips more sharply, and the highest minimum corresponds to the smallest charge. At a certain value of $r_h$, all the temperature curves converge, indicating a universal behavior independent of $Q$. Moreover, we observe that for a higher value of $H(\varepsilon)$, the minimum of $T_H$ shifts towards larger values of $r_h$, and the minimum itself becomes lower.

From this behavior, we understand that the black hole undergoes a phase transition, from an unstable phase where $\frac{\partial T_H}{\partial r_h} < 0$ to a stable one where $\frac{\partial T_H}{\partial r_h} > 0$. Therefore, both the charge $Q$ and the rainbow function $H(\varepsilon)$ influence the conditions for phase transitions and the  thermal stability. We also remark  that the temperature exhibits a singular behavior at \( r_{\text{div}} = \frac{1}{A} \), which is entirely determined by the acceleration parameter. This divergence corresponds to a non-physical point that must be excluded in order to study the physical behavior of both the temperature and the entropy.

\subsection{Stability behaviors}
\begin{figure}[h!]
    \centering
    \begin{tabular}{cc}
       \includegraphics[width=7cm,height=7cm]{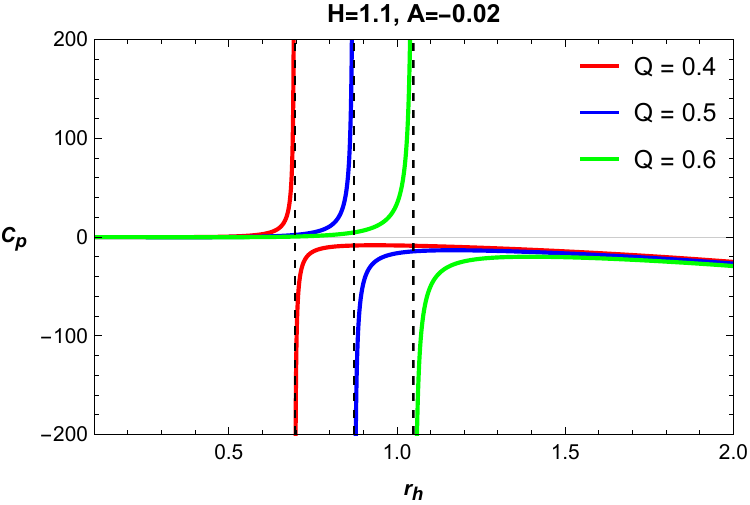} & 
         \includegraphics[width=7cm,height=7cm]{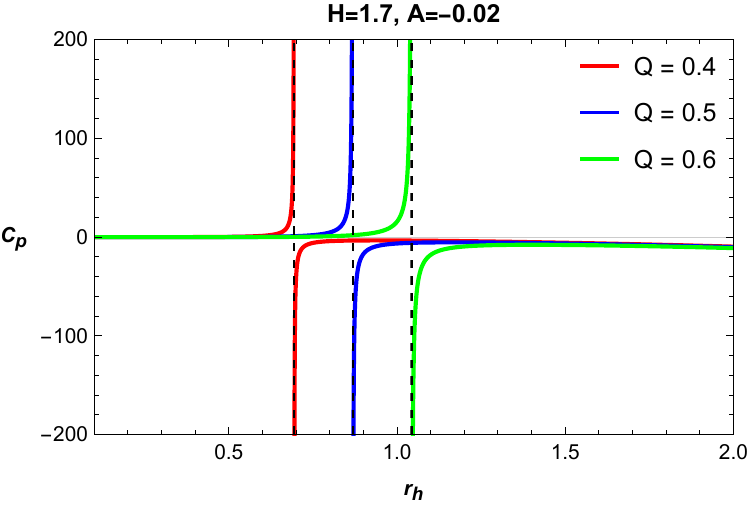} 
    \end{tabular}
    \caption{Effect of the charge parameter $Q$ and the rainbow function $H(\varepsilon)$ on the heat capacity.}
    \label{4}
\end{figure}

To study the local thermodynamic stability of such black holes, we need to  calculate the heat capacity  $C_p$  via the relation
\begin{equation}
C_p = T_H  \frac{\partial S}{\partial T_H},
\end{equation}
where one has used  $S = \frac{A}{4}$  being the entropy of the black hole.  Here,  $A$  denotes the area of the event horizon which reads as 
\begin{equation}
A = \iint \sqrt{g_{\theta\theta} g_{\phi\phi}} \, d\theta \, d\phi \big|_{r=r_h} =  \frac{4\pi r_h^2}{H^2 (\varepsilon)(1 - A^2 r_h^2)K}.   
\end{equation}
Substituting this into the expression of the entropy, we obtain
\begin{equation}
S = \frac{\pi r_h^2}{H^2 (\varepsilon)(1 - A^2 r_h^2)K}.\label{yyy}
\end{equation}
It is important to note that in the absence of acceleration parameters ($K = 1$ and $A = 0$), and in  the standard gravity, the entropy reduces to the familiar form
\begin{equation}
S = \pi r_h^2.
\end{equation}
Using the expression of  $T_H$ given in Eq.(\ref{xxx}) along with the entropy from Eq.(\ref{yyy}), and after carrying out the necessary calculations, the heat capacity can be expressed in the following form
{\footnotesize
\begin{equation}
C_p = \frac{2 \pi r_h^2 \left[ \Lambda (A^2 r_h^2 - 3) r_h^4 - 3H^2(\varepsilon)(Q^2 - r_h^2)(A^2 r_h^2 - 1)^2 \right]}%
{K H^2(\varepsilon)(A^2 r_h^2 - 1) \left[ 3H^2(\varepsilon)(A^2 r_h^2 - 1)^2 (A^2 r_h^4 + r_h^2 + Q^2 (A^2 r_h^2 - 3)) + \Lambda (A^4 r_h^4 + 3) r_h^4 \right]}.
\end{equation}}
In Fig.(\ref{4}), we plot the behavior of  the heat capacity \( C_p \) for different values of \( Q \) and two values of \( H(\varepsilon) \) (1.1 on the left and 1.7 on the right). The results show that increasing the charge \( Q \) leads to an increase in the critical radius \( r_h \), where \( C_p \) diverges, marking the transition between thermodynamically stable and unstable phases. However, we observe that changing the rainbow function \( H(\varepsilon) \) does not affect the position of this critical point, suggesting that the rainbow function has no impact on the local thermodynamic stability of the black hole. This result indicates the presence of a phase transition, which will be further discussed in the context of  Gibbs free energy computations. Furthermore, the analysis reveals that increasing the absolute value of the cosmological constant \( |\Lambda| \) tends to enhance the local stability of the black hole, expanding the range of radii where the heat capacity is positive. A similar effect is observed when increasing the acceleration parameter \( A \), which also leads to a broader locally stable region.  In addition,   it has been  found that variations in the rainbow function \( F(\varepsilon) \) have no noticeable effect on the heat capacity, indicating that the local thermodynamical behavior is independent of this parameter contribution.

\section{$P$-$v$ criticality  and  phase transition }
In this section, we   investigate    the   critical behaviors and the phase transitions  of such charged black hole solutions by computing the associated thermodynamical  quantities.
 \subsection{$P$-$v$ criticality} 
\begin{figure}[h!]
    \centering
    \begin{tabular}{cc}
       \includegraphics[width=7cm,height=7cm]{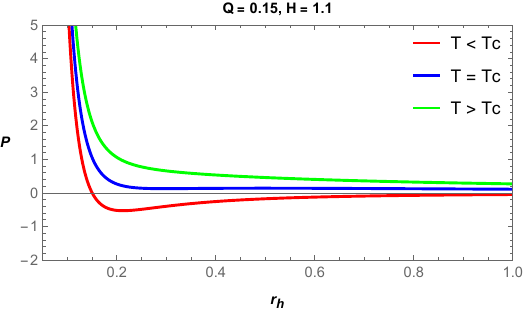} & 
         \includegraphics[width=7cm,height=7cm]{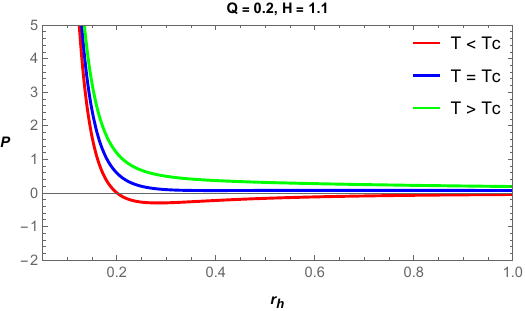} 
    \end{tabular}
\caption{Pressure of charged AdS black holes in rainbow gravity as a function of the horizon radius for different temperature values. }
\label{F41}
\end{figure}

To   discuss  the $P$-$v$  criticality behaviors of such charged AdS  black hole solutions, we  should  determine the thermodynamic state equation. This can be established   by   considering 
the cosmological constant $\Lambda $   as a pressure\begin{equation}
P=-\frac{\Lambda }{8\pi }.
\end{equation}%
After computations,  we find the expression of   the pressure 
\begin{equation}
 P = \frac{ -3H(\varepsilon) \left( A^2 r_h^2 - 1 \right) \left( 4\pi T_H F(\varepsilon) r_h^3 + H(\varepsilon) \left( r_h^2 (A^2 Q^2 + 1) - Q^2 - A^2 r_h^4 \right) \right) }{ 8\pi r_h^4 \left( A^2 r_h^2 - 3 \right) }.   \label{P}
\end{equation}
 We identify the black hole thermodynamic volume as 
 \begin{equation}
 V=\dfrac{\partial M}{\partial P}=\dfrac{4 \pi r_{h}^3}{3(1-A^2 r_{h}^2)H^2(\varepsilon)}.\label{V}
 \end{equation}
To obtain  of  the critical pressure  $P_c$,  the critical specific volume $r_c$
and the  critical temperature $T_c$, we follow the method used in \cite{12}. According to such a work, we consider  $Ar_h$ as a constant   $a$  being different to  one as follows
\begin{equation} 
Ar_h=a.
\end{equation}
In this way,  the  the critical points can determined  from  the equations
\begin{equation}
\frac{\partial P}{\partial r_h }=0,\hspace{1.5cm}\frac{\partial ^{2}P}{%
\partial r_h^{2}}=0.
\end{equation}
Performing  the computations, the critical quantities are found to be 
\begin{eqnarray}
P_c&=&\frac{H^{2}(\varepsilon) \left(a^{2}-1\right)}{32 Q^{2} \left(a^{2}-3\right) \pi} \nonumber\\ 
T_{c}&=&\frac{H(\varepsilon) \sqrt{6}}{18 Q \pi  F(\varepsilon)} \nonumber\\
r_{c}&=&\sqrt{6}\, Q , \nonumber
\end{eqnarray}
where an extra  constraint should be  imposed on the constant $a$ being $a\neq \sqrt{3}$.
  To this end, the critical triple $(P_{c}, T_{c}, r_{c})$ yields the following ratio
\begin{equation}
\chi=\dfrac{P_{c}r_{c}}{T_{c}}=\frac{9 H(\varepsilon) \left(a^{2}-1\right) F(\varepsilon)}{16 (a^{2}-3)},
\end{equation}
which  does not retain a constant value as  charged  AdS black holes \cite{17}.  However,  this expression  can recover certain known relations.  Taking, for instance $F(\varepsilon)=H(\varepsilon)=1$, we obtain  the ratio
\begin{equation}
\chi=\dfrac{P_{c}r_{c}}{T_{c}}=\frac{9  \left(a^{2}-1\right) }{16 (a^{2}-3)},
\end{equation}
reported in \cite{12}. In the small parameter limits,  the behavior of the extended universal ratio  with  $ v_c=2r_c$  takes the form
\begin{equation}
\dfrac{P_{c}v_{c}}{T_{c}}=\dfrac{3}{8}+ \dfrac{\varepsilon^2}{8}+  \dfrac{a^2(\varepsilon^2-2)}{8}+O(\varepsilon^3, a^3),
\end{equation}
 which is  a universal behavior with respect to the electric charge $Q$. Taking  $a =  \varepsilon= 0$,  we  recover  the  RN-AdS black hole situation  \cite{17,177,1777}.  
In Fig.(\ref{F41}), we plot the $P-v$ diagram by considering the horizon  radius $r_h$ instead of the specific volume $v$. To illustrate the effect of the electric charge on the presence of rainbow gravity, we fix the value of $H$ and consider two different values of the charge. It is evident from the figure that for $T<T_c$, the black hole system exhibits behavior similar to that of an extended Van der Waals gas, with inflection points indicating critical behavior. This critical behavior persists in the presence of rainbow gravity and continues to depend on the value of the electric charge.

\subsection{Joule-Thompson expansion}
To get more information on the proposed black hole thermodynamics, we approach   the Joule-Thompson expansion \cite{1000,2000}.  Fixing the charge,  the Joule-Thomson coefficient can be expressed as follows 
\begin{equation}
\mu=\left( \dfrac{\partial T}{\partial P} \right)_{M}=\dfrac{1}{C_{P}} \left[ T \left( \dfrac{\partial V}{\partial T} \right)_{P}-V  \right] \label{m}
\end{equation}
For later use,  the equation of state for such a black hole  should be given  terms of the thermodynamic volume.
Considering Eq.(\ref{V}), Eq.(\ref{P}) and Eq.(\ref{xxx}), we get  the temperature as a function  of the volume and the pressure
\begin{equation}
T = \dfrac{1}{12 V (a^2-1)F(\varepsilon) H(\varepsilon)} \left(  
3 P V (3-a^2){\cal V}^{\!1/3} +(a^2-1){\cal V}^{\!2/3} - 4Q^2  \right)  \label{t}
\end{equation}
where one has used ${\cal V}=\dfrac{6 V(1-a^2)H^2 (\varepsilon)}{\pi}$.  Using Eq.(\ref{t}) and the second part of Eq.(\ref{m}), we can find the temperature associated with a zero Joule-Thomson coefficient.  Indeed, the  repeated inversion temperature $T_i$ is found to be 
{\footnotesize \begin{equation}
T_i = \dfrac{1}{36 \pi F(\varepsilon) H(\varepsilon)(1-a^2)^{2/3}} \left(  
8P\pi(3-a^2)V^{1/3} + \dfrac{3H^2(\varepsilon)}{V} (1-a^2)^{2/3} \left(3Q^2 - ((1-a^2)V)^{2/3}\right) 
\right)
\end{equation}}
Introducing the volume quantity,  this temperature  can take the following form 
\begin{equation}
T_i= \frac{9 \left(a^2 -1\right)^{2} \left(Q^{2}+\frac{r^{2}}{3}\right)  H^{2}(\varepsilon)-8 P \pi \,r^{4}   \left(a^{2}-3\right)}{36 \pi  F(\varepsilon) H(\varepsilon)  \left(a^{2}-1\right) r^{3}}.
\label{ti1}
\end{equation}
Using  Eq. (\ref{t}), one can get also 
\begin{equation}
T=\frac{3 \left(a^{2}-1\right)^{2} \left(Q^{2} -r^2\right) H^{2}(\varepsilon) -8   P \pi \,r^{4} \left(3-a^{2}\right)}{12 \pi F(\varepsilon)  H(\varepsilon)  \left(a^{2}-1\right)r^{3} }.\label{ti2}
\end{equation}
Subtracting Eq. (\ref{ti1}) form Eq. (\ref{ti2}),  we find the algebraic equation
 \begin{equation}
\pi  P_{i} \,r^{4}-9 H^{2}(\varepsilon) D Q^{2}+6 D H^{2} (\varepsilon) r^{2}=0,
 \end{equation}
where   one used $D=\dfrac{(1-a^2)^2}{(3-a^2)}$   and where $P_{i}$ is the inversion pressure.  Solving this equation, we can get  four roots. However, only one of them has physical significance, while  the others are either complex or negative which  should be evinced.  We focus only  on the real and positive root given by 
 \begin{equation}
r = \sqrt{3} \,
\sqrt{
  H(\varepsilon) \sqrt{
    \frac{D \left(P_{i} H(\varepsilon) Q^{2} + D H^{2}(\varepsilon)\right)}{\pi P_{i}}
  }
  - \frac{D H^{2}(\varepsilon)}{\pi P_{i}}
}.
\end{equation}
At zero inversion pressure $P_i=0$, the inversion temperature reaches its minimum value
 \begin{equation}
T_{i}^{min}=-\frac{ \sqrt{6} \left(a^{2}-1\right) H(\varepsilon) }{36 Q F(\varepsilon) \pi}. 
 \end{equation}
This provides a  ratio between minimum inversion and critical temperatures being 
 \begin{equation}
\xi= \dfrac{T_{i}^{min}}{T_{c}}=\dfrac{1}{2}-\dfrac{a^2}{2}.
 \end{equation}
 In the limit $a=0$, we find that the result $\xi=\dfrac{1}{2}$ reported  in \cite{2000}.  This reveals that  the  obtained  result perfectly with the charged AdS black hole universal behaviors with respect to the charge. 
\subsection{Phase transitions}

\begin{figure}[h!]
    \centering
    \begin{tabular}{cc}
       \includegraphics[width=7cm,height=7cm]{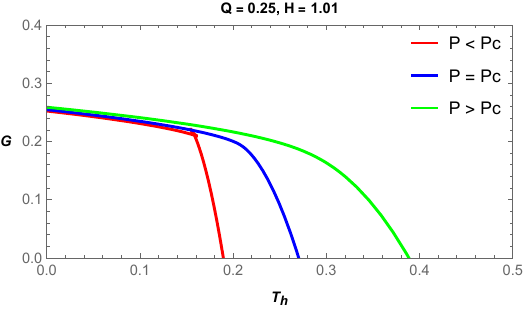} & 
         \includegraphics[width=7cm,height=7cm]{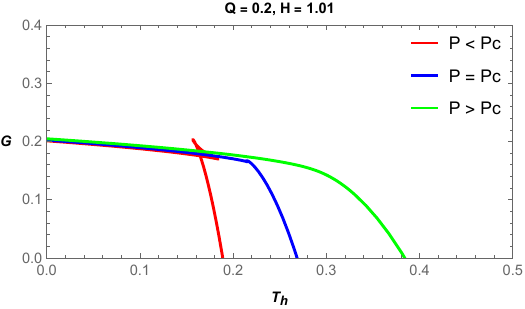} 
    \end{tabular}
    \begin{tabular}{cc}
       \includegraphics[width=7cm,height=7cm]{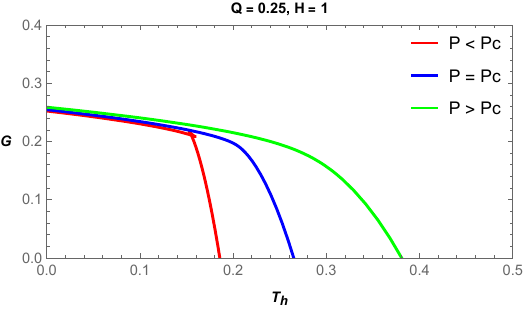} & 
         \includegraphics[width=7cm,height=7cm]{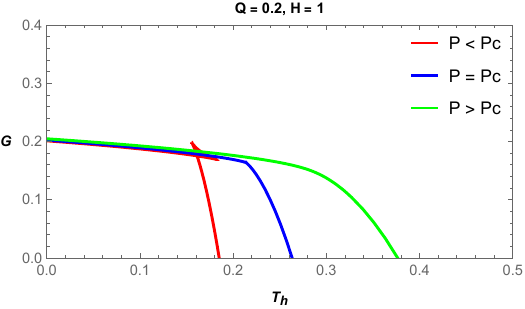} 
    \end{tabular}
    \caption{Gibbs free energy in terms of the temperature for different values of $P$ and $Q$.}
    \label{7}
\end{figure}

To examine  the phase transitions, we  calculate  the Gibbs free energy via the relation  
\begin{equation}
   G = M - T_H S.
\end{equation}
The computations give
{\footnotesize
\begin{equation}
G = \frac{
    -8\pi P   \left(3 - A^2 r_h^2 + 2K F B\right) r_h^4
    + 3 B^2 \left[
        (Q^2 - r_h^2) + 2K F(\varepsilon) (Q^2 + r_h^2) H(\varepsilon)
    \right]
}{
    12K r_h B^2 F(\varepsilon) H(\varepsilon)
},
\end{equation}
}
where one has used \( B = (A^2 r_h^2 - 1) H(\varepsilon) \). Using the critical thermodynamical quantities, the $G-T_H$ curves are given in Fig(\ref{7}).

 This figure illustrates the combined effect of the charge parameter $Q$ and the rainbow function $H(\varepsilon)$ on the thermodynamic phase transitions of the black hole. It is noteworthy that the $G$–$T_H$ (Gibbs free energy versus temperature) curves exhibit similar qualitative behavior for various values of the critical pressure $P_c$. Specifically, for pressures below the critical value ($P < P_c$), the characteristic swallow-tail structure appears in the $G$–$T_H$ diagram. This feature is a hallmark of first-order phase transitions, signifying a transition between small and large black hole phases. This behavior is very similar to that of Van der Waals fluids.  The effect of the electric charge on this behavior is modulated by the presence of rainbow gravity, as encoded in the function $H(\varepsilon)$. The deformation introduced by rainbow gravity influences the location and nature of the phase transition, thereby controlling how the charge parameter $Q$ affects the thermodynamic properties of the black hole.
\\
\vspace{2cm}
\section{Conclusions}
In this work, we   have  studied  the thermodynamical properties  of accelerated charged
AdS black holes in the context of rainbow gravity.   Precisely,   we  have   examined   the  thermal 
stability and   the critical behaviors including the phase transitions.   In particular, we have computed  the relevant theromodynalical quantities. Using the associated laws, we have calculated  the heat  capacity in order to inspect the  stability  behaviors. We have determined the regions  where the  black holes are stable.  Connecting the  acceleration parameter $A$ and the horizon radius $r_h$ via a constant parameter $a$,  we have  approached  
the $P-v$ criticality  by  calculating  the critical pressure
$P_c$,  the critical temperature $T_c$ and the  critical specific volume $r_c$ in terms of  $a$ and  the rainbow gravity parameter  $\varepsilon$.   We have shown   that  the ratio $ \dfrac{P_{c}r_{c}}{T_{c}}$
is a universal number with respect to the charge $Q$.  In small limits of the external parameters,  we  have recovered   the Van der Waals fluid    behaviors. Finally, we have examined  the Joule-Thomson expansion effects  for  such black holes.   We have   revealed  that the  similarities and  the differences with  Van der Waals fluids depend on the region parameters.

This work leaves certain open questions. A natural question is to consider other properties including the optical ones. The shadow and  the deflection angle of lights near to the black holes could be addressed in future works.

  {\bf Data availability}\\
  Data sharing is not applicable to this article.

\section*{Acknowledgements}
The authors  would like to thank S. E. Baddis  and A. Belhaj for discussions and encouragement.  This work was done with the support of the CNRST in the frame of the PhD Associate Scholarship Program PASS.


\begin{thebibliography}{99}

\bibitem{1} A. Einstein, \textit{The Field Equations of Gravitation}, Sitzungsberichte der Preussischen Akademie der Wissenschaften, Berlin (1915).


\bibitem{102}
Z.-Y. Fan and X. Wang,\textit{ Construction of Regular Black Holes in General Relativity,}
 Phys. Rev. D 94 (2016) 124027, \texttt{arXiv:1610.02636}.

\bibitem{2} C. T. Bolton, \textit{Identification of Cygnus X-1 with HDE 226868}, Nature \textbf{235} (1972) 271.

\bibitem{3} B. P. Abbott et al., \textit{Observation of Gravitational Waves from a Binary Black Hole Merger}, Phys. Rev. Lett. \textbf{116} (2016) 061102.



\bibitem{4} K. Akiyama and al., \textit{First M87 Event Horizon Telescope
Results. IV. Imaging the Central Supermassive Black Hole}, Astrophys. J.
\textbf{L4} (1) (2019) 875, \texttt{arXiv:1906.11241}.

\bibitem{401} K. Akiyama and al., \textit{First M87 Event Horizon Telescope
Results. V. Imaging the Central Supermassive Black Hole}, Astrophys. J.
\textbf{L5} (1) (2019) 875.

\bibitem{402} K. Akiyama and al., \textit{First M87 Event Horizon Telescope
Results. VI. Imaging the Central Supermassive Black Hole}, Astrophys. J.
\textbf{L6} (1) (2019) 875.



\bibitem{6} J. M. Bardeen, B. Carter, S. W. Hawking, \textit{The Four Laws of Black Hole Mechanics}, Commun. Math. Phys. \textbf{31} (1973) 161.




\bibitem{71} S. W. Hawking, D. N. Page, \textit{Thermodynamics of black holes
in anti-de Sitter space}, Commun. Math. Phys. \textbf{87} (4) (1983) 577.

\bibitem{72} M. Cvetic and al, \textit{Embedding AdS black holes in ten and
eleven dimensions}, Nucl. Phys. B\textbf{558} (1999) 96, \texttt{
arXiv:hep-th/9903214}.



\bibitem{73} S. W. Wei, Y. C. Zou, Y. X. Liu, R. B. Mann, \textit{Curvature
radius and Kerr black hole shadow}, JCAP \textbf{08} (2019)030, \texttt{%
arXiv:1904.07710}.

\bibitem{74} A. Belhaj, H. Belmahi, M. Benali, W. El Hadri, H. El Moumni, E.
Torrente-Lujan, \textit{Shadows of 5D Black Holes from string theory}, Phys.
Lett. B\textbf{812} (2021)136025, \texttt{arXiv:2008.13478}.

\bibitem{75} A. Belhaj, H. Belmahi, M. Benali, \textit{Superentropic AdS
black hole shadows}, Phys. Lett. B \textbf{821} (2021)136619, \texttt{%
arXiv:2110.06771}.
\bibitem{76}  H. Belmahi,  \textit{ Constrained Deflection Angle and Shadows of Rotating Black Holes in Einstein-Maxwell-scalar Theory},  \texttt{ arXiv:2411.11622}.
\bibitem{77} 
D. J. Gogoi and S. Ponglertsakul,   \textit{ Constraints on quasinormal modes from black hole
shadows in regular non-minimal Einstein Yang–Mills gravity}, Eur. Phys. J. C 84 (2024) 652.

\bibitem{78} P. V. P. Cunha, C.A. R. Herdeiro, B. Kleihaus, J.Kunz, E. Radu,
\textit{\ Shadows of Einstein-dilaton-Gauss-Bonnet black holes}, Phys. Lett.
B \textbf{768}(2017) 773, \texttt{arXiv:1701.00079}.

\bibitem{79}
D.~Kubiznak, R.~B.~Mann and M.~Teo,
\textit{Black hole chemistry: thermodynamics with Lambda},
Class. Quant. Grav. \textbf{34} (2017) 063001, \texttt{arXiv:1608.06147}.


\bibitem{80} Y. Liu, D. C. Zou, B. Wang, \textit{Signature of the Van der
Waals like small-large charged AdS black hole phase transition in quasi
normal modes}, JHEP \textbf{09} (2014) 179, \texttt{arXiv:1405.2644}.

\bibitem{81} A. Belhaj, A. El Balali, W. El Hadri, E. Torrente-Lujan,
\textit{On Universal Constants of AdS Black Holes from Hawking-Page Phase
Transition}, Phys. Lett. B\textbf{811} (2020) 135871, \texttt{%
arXiv:2010.07837}.

\bibitem{82} A. Rajagopal, D. Kubiznak, R. B. Mann, \textit{Van der Waals
black hole}, Phys. Lett. B\textbf{737} (2014) 277, \texttt{arXiv:1408.1105}.

\bibitem{83} A. Belhaj, H. Belmahi, M. Benali, A. Segui, \textit{
Thermodynamics of AdS black holes from deflection angle formalism}, Phys.
Lett. B\textbf{817} (2021) 136313.

\bibitem{Karch}
A.~Karch and B.~Robinson,
 \textit{Holographic Black Hole Chemistry,}
JHEP \textbf{12} (2015).
\bibitem{7} S. W. Hawking, \textit{Particle Creation by Black Holes}, Commun. Math. Phys. \textbf{43} (1975) 199.

\bibitem{8} J. D. Bekenstein, \textit{Black Holes and Entropy}, Phys. Rev. D \textbf{7} (1973) 2333.

\bibitem{9} S. W. Hawking, D. N. Page, \textit{Thermodynamics of Black Holes in Anti-de Sitter Space}, Commun. Math. Phys. \textbf{87} (1983) 577.
\bibitem{m1}
R.~Li, K.~Zhang, J.~Yang, R.~B.~Mann and J.~Wang,
 \textit{Critical slowing down of black hole phase transition and kinetic crossover in supercritical regime}, \texttt{[arXiv:2505.24148 ]}.
\bibitem{m3}
J.~Yang and R.~B.~Mann,
\textit{Dynamic behaviours of black hole phase transitions near quadruple points,}
JHEP \textbf{08} (2023), 028
doi:10.1007/JHEP08(2023)028.
 \bibitem{99}
H. El Moumni and K. Masmar, Regular AdS black holes holographic heat engines in a
benchmarking scheme, Nucl. Phys. B {\bf  973} (2021), 115590.

\bibitem{Mancilla}
R.~Mancilla,
 \textit{Generalized Euler Equation from Effective Action: Implications for the Smarr Formula in AdS Black Holes,},
\texttt{arXiv:2410.06605}.
 \bibitem{999} K. Masmar,  \textit{
Joule–Thomson expansion for a nonlinearly charged Anti-de Sitter black hole},  Int. J.Geom.Meth.Mod.Phys. {\bf 20} (2023) 05, 2350080.
\bibitem{EslamPanah:2025zcm}
B.~Eslam Panah,
\textit{Super-entropy bumblebee AdS black holes,}
Phys. Lett. B \textbf{861} (2025).
\bibitem{10} E. Witten, \textit{Anti-de Sitter Space and Holography}, Adv. Theor. Math. Phys. \textbf{2} (1998) 253.

\bibitem{11} A. Anabalon et al., \textit{Accelerated Black Holes beyond the C-metric}, JHEP \textbf{04} (2018) 096.
 \bibitem{111} L. Chakhchi, H. El Moumni, K. Masmar,  \textit{
A Supersymmetric Suspicion From Accelerating Black Hole Shadows},  \texttt{ arXiv:2409.02594}.



\bibitem{12} H. Liu, X.-h. Meng, \textit{$P-V$ Criticality In the Extended Phase Space of Charged Accelerating AdS Black Holes}, 	Mod. Phys. Lett. A, Vol. 31, No. 37 (2016) 1650199, \texttt{arXiv:1607.00496v2}
\bibitem{m2}
T.~Hale, D.~Kubiz{\v{n}}{\'a}k, J.~Men{\v{s}}{\'\i}kov{\'a}, R.~B.~Mann and J.~Yang,
\textit{Thermodynamics of charged and accelerating black holes,}
Phys. Rev. D \textbf{111} (2025) no.10, 104004.

\bibitem{13}
J. Magueijo, L. Smolin, \textit{Gravity’s Rainbow}, 	Class.Quant.Grav. 21 (2004) 1725-1736, \texttt{arXiv:gr-qc/0305055v2}
\bibitem{EslamPanah}
B.~Eslam Panah, N.~Heidari, M.~Soleimani and M.~Kaveh, \textit{
Super-entropic black holes in gravity{\textquoteright}s rainbow and determining constraints on rainbow functions,}
Eur. Phys. J. C \textbf{85} (2025)
\bibitem{14} A. Anabalon et al., \textit{Accelerated Black Holes beyond the C-metric}, JHEP \textbf{04} (2018) 096.

\bibitem{15}
M. Appels, R. Gregory, D. Kubiznak, \textit{Thermodynamics of Accelerating Black Holes}, Phys. Rev. Lett. 117, 131303 (2016), \texttt{arXiv:1604.08812}
\bibitem{EslamPanah:2024dfq}
B.~Eslam Panah, S.~Zare and H.~Hassanabadi,
\textit{Accelerating AdS black holes in gravity{\textquoteright}s rainbow,}
Eur. Phys. J. C \textbf{84}   259 (2024), \texttt{arXiv:2403.09757
}.
\bibitem{16}
A. Anabalon, M. Appels, R. Gregory, D. Kubiznak, R. B. Mann, A. Övgün, \textit{Holographic Thermodynamics of Accelerating Black Holes}, 	Phys. Rev. D 98, 104038 (2018), \texttt{arXiv:1805.02687}

\bibitem{17}
D. Kubiznak, R. B. Mann, \textit{P-V criticality of charged AdS black holes}, JHEP 1207:033,2012, \texttt{arXiv:1205.0559}
\bibitem{177}
A. Belhaj, M. Chabab, H. El Moumni and M. B. Sedra,  \textit{ On Thermodynamics of
AdS Black Holes in Arbitrary Dimensions}, Chin. Phys. Lett. 29 (2012) 100401,
arXiv:1210.4617 [hep-th].
\bibitem{1777}
A. Belhaj, M. Chabab, H. El Moumni, K. Masmar, M. B. Sedra and A. Segui,  \textit{ On
Heat Properties of AdS Black Holes in Higher Dimensions}, JHEP 05 (2015), 149, arXiv:1503.07308 [hep-th].

\bibitem{1000}
M. Rostami, J. Sadeghi, S. Mirabotalebi, A. A. Masoudi, B. Pourhassan,\textit{ Charged accelerating AdS black hole of f R gravity and the Joule
Thomson expansion}, Int. J. Geom. Methods Mod. Phys(2020)

\bibitem{2000}
Ö. Ökcü, E. Aydıner, \textit{Joule-Thomson Expansion of Charged AdS Black Holes}, 	Eur. Phys. J. C (2017) 77, \texttt{arXiv:1611.06327}
\end{thebibliography}
\end{document}